\documentclass[mathleft
]{an}
\usepackage{graphicx}
\usepackage{times}
\usepackage{amssymb}
\usepackage[normal]{subfigure}
\usepackage[fleqn]{amsmath}

\begin{document}

\Pagespan{1}{}
\Yearpublication{}%
\Yearsubmission{}%
\Month{}%
\Volume{}%
\Issue{}%

\title{Habitability and Multistability in Earth-like Planets}

\author{Valerio Lucarini\inst{1,2}\fnmsep\thanks{Corresponding author:
  \email{valerio.lucarini@zmaw.de}\newline}
\and  Salvatore Pascale \inst{1} \and Robert Boschi \inst{1} \and Edilbert Kirk \inst{1} \and Nicolas Iro\inst{1}
}
\titlerunning{}
\authorrunning{Lucarini \& al.}
\institute{
Klimacampus, Meteorologisches Institut, Universit\"at  Hamburg, 
Hamburg, Germany
\and 
Department of Mathematics and Statistics, University of Reading, Reading, UK}

\received{}
\accepted{}
\publonline{later}


\abstract{In this paper we explore the potential multistability  of the climate for a planet around the habitable zone. We focus on conditions reminiscent to those of the Earth system, but our investigation has more general relevance and aims at presenting a general methodology for dealing with exoplanets. We describe a formalism able to provide a thorough analysis of the non-equilibrium thermodynamical properties of the climate system and explore, using a a flexible climate model, how such properties depend on the energy input of the parent star, on the infrared atmospheric opacity, and on the rotation rate of the planet. We first show that it is possible to reproduce the multi-stability properties reminiscent of the paleoclimatologically relevant snowball (SB) - warm (W)  conditions. We then characterise the thermodynamics of the simulated W and SB states, clarifying the central role of the hydrological cycle in shaping the irreversibility and the efficiency of the W states, and emphasizing the extreme diversity of the SB states, where dry conditions are realized. Thermodynamics provides the clue for studying the tipping points of the system and leads us to constructing empirical parametrizations allowing for expressing the main thermodynamic properties as functions of the emission temperature of the planet only. Such empirical functions are shown to be rather robust with respect to changing the rotation rate of the planet from the current terrestrial one to half of it. Furthermore, we explore the dynamical range of slowy rotating and phase locked planet, where the length of the day and the length of the year are comparable. We clearly find that there is a critical rotation rate below which the multi-stability properties are lost, and the ice-albedo feedback responsible for the presence of SB and W conditions is damped. The bifurcation graph of the system suggests the presence of a phase transition in the planetary system. Such critical rotation rate corresponds roughly to the phase lock 2:1 condition. Therefore, if an Earth-like planet is 1:1 phase locked with respect to the parent star, only one climatic state would be compatible with a given set of astronomical and astrophysical parameters. These results have relevance for the general theory of planetary circulation and for the definition of necessary and sufficient conditions for habitability.}

\maketitle

\section{Introduction}

The investigations of extra-solar planetary objects is one of the most active fields of research in astrophysical sciences. After the first discoveries dating back to the mid 1990s, improved observational datasets and techniques of data analysis have made possible to catalogue and studies hundreds of planets orbiting parent stars other than the Sun. These planets come in a great variety of physical, chemical, and orbital properties. Factors of great relevance include the presence or lack of a rocky core, the composition of the atmosphere, the color intensity of the radiation emitted by the parent star, the orbital parameters, the presence or lack of a tidal-lock condition. An especially important problem for rocky planets (the so-called super-Earths) is determining under which conditions one can expect that the planet may feature the presence of liquid water at surface, as this is seen as necessary condition for the presence of life. We refer the reader to some recently published  books \nocite{Dvorak,Seager, Kasting, Perryman} (Dvorak 2008; Kastings 2009; Seager 2010; Perryman 2011) .

The existence of great variety of planetary conditions is, on one side, obviously a scientific gold mine, and on the other side it is a maze one can easily get lost in. It is incredibly attractive to have the possibility of studying the individual properties of a fast growing number of newly-discovered astronomical objects, for which a growing amount of information is coming and is very likely to come in the near and medium-term future, considering the technological development and the coming astrophysical big science initiatives. On the other side, one may demand what is in the long term the scientific merit of performing a \textit{Linnaean} compilation of the properties of distant planets, and of developing radiative models and/or general circulation models aimed at describing the radiative and dynamical properties of their atmospheres. We know that, even for planets we know well, like those of our solar system, developing accurate numerical models able to account for the dynamics and thermodynamics of the planetary atmosphere compatibly with our observations is extremely challenging. And, coming back to Earth, it is well-known that there is no satisfactory theory of climate dynamics, able to account for the complex, multi-scale interplay between radiative forcings, dynamical instabilities, positive and negative feedbacks, impacts of slow modulations of orbital and geological conditions. The development of a complete climate theory is, in fact, one of the great challenges of contemporary science, and the development of seamless models able to account for the behavior of the fluid component of our own planet over a large range of temporal and spatial scale is still a distant hope. The theoretical interpretation of the onset and decay of many well-documented events of our planet's history - like the ice ages - is far from settled,  let alone the presence of models able to simulate them satisfactorily. A provocative question one may formulate using a reductionistic point of view may well be: what is the hope/goal/scope of studying the properties of the atmospheres of distant planets if we still know relatively little of the properties of our own? 

We argue here that, in fact, the investigation of exoplanets creates an outstanding scientific horizon for transdisciplinary research across astrophysics and geophysics, and, in fact, geophysical sciences can definitely contribute to the investigation of non-terrestrial objects, and, on the other side, by studying other planets, we can understand better the planet Earth, because we can take a broader scientific perspective, less centered on trying to predict and explain the variability of some specific physical properties of the Earth's fluids (typically of direct interest for mankind) preferentially over other ones. The challenge is then trying to develop general physical and mathematical theoretical  methods and diagnostic tools to be used for studying a large variety of planetary atmosphere, and, at a more practical level, to construct flexible numerical models able to simulate many different planetary atmospheres once some basic input parameters are given. In this perspective, one must refer to recent paper by  \cite{Read}, where, along the lines of what traditionally done in fluid mechanics, it is suggested to construct several dimensionless numbers  and use them for classifying the circulation regimes of classes of planets. 

In this paper we try to give a limited, temptative but three-fold contribution to establishing such a link between geophysical and astrophysical sciences. We have drawn from some recently published material \nocite{Lucarini,GenSens,LucFr,Luc10,Bob} (Lucarini 2009; Lucarini, Fraedrich \& Lunkeit 2010a; 2010b; Boschi et al. 2013) and we add some new results of distinct astronomical flavor. First, we introduce some ideas and methods derived from basic results of macroscopic thermodynamics that allow to define in a rigorous way indicators describing the basic non-equilibrium  properties of a general planetary atmosphere: its ability to transform available potential energy into kinetic energy, thus performing work like a thermal engine; and its production of entropy through irreversible process; and how these two concepts can be related in a compact conceptual framework. Then, we use the derived thermodynamical indicators to provide a novel description of a classical problem of paleoclimatology, \textit{i.e.} the onset and decay of the so-called \textit{Snowball Earth}. Based on the evidence supported by \cite{Hoffman} and \cite{HoffmanSchrag}, it is  expected that the Earth is potentially capable of supporting multiple steady states for the same values of some parameters such as  the solar constant and the concentration of carbon dioxide, which directly affect the radiative forcing. Such states are the presently observed state, and a state where the planet is entirely ice covered and the surface temperature is much lower than the present one. Such conditions hardly allow for the presence of life, so this issue is of extreme relevance of the general quest for defining habitability condition in other planets.  Initial research using simple $0$-D models \nocite{Budyko,Sellers} (Budyko 1969; Sellers 1969), $1$-D models \nocite{Ghil} (Ghil 1976) as well as more recent analyses performed using complex 3-D general circulation models \nocite{Maro,Voigt,Pierre11} (Marotzke \& Botztet 2007; Voigt \& Marotzke 2011; Pierrehumbert et al. 2011),  provide support for the existence of such bistability for certain range of parameters of the system. The main mechanism triggering the abrupt transitions between the snowball and the warm state is the positive ice-albedo feedback  \nocite{Budyko,Sellers} (Budyko 1968; Sellers 1969). Such a feedback is associated with the fact that as temperatures increase, the extent of snow and ice cover decreases thus reducing the albedo and therefore increasing the amount of sunlight absorbed by the Earth system. Conversely, a negative fluctuation in the temperature  leads to an increase in the albedo therefore reinforcing the cooling. 
We show how thermodynamics allows for a much deeper physical interpretation of the ice-albedo feedback, which is the instability mechanism mainly responsible for the multi stability of the system. At this regard, we use the numerical evidence gathered by running PlaSim, a general circulation model of intermediate complexity \cite{Frae2}, and study some basic properties of the climate states realized when the solar constant is modulated between $1160$ Wm$^{-2}$  and $1510$ Wm$^{-2}$ and the  values of [CO$_2$] are varied between  90 and 2880 ppm, and define the region of multistability. Then, we extend the previous result in the direction of providing information useful for general planetary objects by altering a critical astronomical parameter, \textit{i.e.} the rotation rate of the planet, and explore the case of slowly rotating planets. We perform these investigation with a modified climate model, where, for sake of generality, the land-sea contrast typical of our planet is removed and an ocean only surface is considered. Whereas for the climate of a so-called \textit{Aquaplanet} is indeed different from that of the corresponding realistic Earth, its structural properties in terms of multi stability are the same \cite{Voigt2}, so we gain generality without losing comparability of the results. We then study how the structural properties of the system change when we change the rotation rate of the planet. We discover that when the rotation rate become smaller than a critical but non-zero value, the multi stability condition is lost, and, in particular, the 1:1 phase locked planet has a unique domain of attraction for its physical conditions. The exploration of very diverse dynamical regimes determined by the rotation rate is performed using  a prototypical version of a flexible modeling suite which is being built starting from the (already rather flexible) PLASIM platform. Such modeling suite, once completed, will allow for simulating extremely different super-Earths, using exactly the same dynamical core and modeling platform. 

The paper is organized as follows. In Section \ref{Noneq} we provide a basic introduction to some tools of non-equilibrium thermodynamics used throughout the paper. In Section \ref{Exp} we describe our numerical simulations and the set-up of the model used in this study. In Section \ref{Results} we discuss the bistability properties of the climate state, discussing the Snowball and the Snow-free states, and the transitions between the two. in Section \ref{rotation} we present the results obtained on the role of the rotation rate in defining the structural properties of the system. In Section \ref{concl} we draw our conlusions.

\section{Non-equilibrium Thermodynamics of the climate}
\label{Noneq}

In this section we  briefly introduce the  thermodynamic diagnostic tools  used in this review to analyse the multistability properties of an Earth-like  planet. An extended  discussion of such a  formalism and of its applications to study global properties of non-equilibrium systems  can be found elsewhere \nocite{Frae, Lucarini,GenSens,LucFr} (Fraedrich \& Lunkeit 2008; Lucarini 2009; Lucarini et al. 2010; 2010b). 

The total energy budget of the atmosphere can be written as $E = P + K$, where $K=\int_V\rho \mathbf{v}^2/2 dV$ represents the total kinetic energy and $P=\int_V (c_pT+gz+Lq)\rho dV$ is the moist static potential energy (which is the sum of  sensible,  potential and latent energy) and   $V$ is the atmospheric domain. It can be shown \nocite{Peix2} (Peixoto \& Oort 1992) that:
\begin{align}
   \dot{K}=-D+W    \label{kin}, \\ 
  \dot{P}=\dot{\Psi}+D-W,  \label{pot}
  \end{align}
    where $ D=\int_{V}dV \rho\epsilon^2$  is the dissipation of kinetic energy due to friction ($\epsilon^2 >0$ is the local rate of heating associated with kinetic energy  dissipation), $W=-\int_V \mathbf v\cdot\nabla p\rho dV $ is the instantaneous work done by the system and $\dot{\Psi}=\int_{V}d V (-\nabla\cdot H)$ is the total heating due to convergence of sensible heat, latent heat   and radiative fluxes. Equations (\ref{kin})-(\ref{pot}) imply that  $\dot{E}=\dot{\Psi}$ and therefore  the frictional heating $\epsilon^2$ does not increase the total energy since  it is just an internal conversion between kinetic and potential energy.   Taking the climate as a non-equilibrium steady state system (NESS, see \cite{Gallavotti}), we have that over long time scales $\overline{\dot{E}}=\overline{\dot{P}}=\overline{\dot{K}}=0$  (here and in the following the  bar indicates averaging over long time periods) and therefore $\overline{\dot{\Psi}}=0$.   If we  define the total diabatic heating  $\dot{Q}=\rho \epsilon^2 -\nabla \cdot H$ and split the domain $V$ into the subdomain $V^+$ , where $ \dot{Q}=\dot{Q}^+ >0$,   and $V^-$,   where $\dot{Q}=\dot{Q}^- <0$,  it can be easily seen from Eq. (\ref{pot}) that:
\[
\dot{\Psi}+D=\dot{P}+W=\int_{V^+} dV \rho Q^+  + \int_{V^-} dV \rho  Q^- = 
\]
\begin{equation} 
\label{phipm}
=\dot{\Phi}^+ + \dot{\Phi}^-,
\end{equation}
with $\overline{\dot{\Phi}^+}>0$ and $\overline{\dot{\Phi}^-}<0$ by definition.
From eq. (\ref{kin}-\ref{pot}) and (\ref{phipm})  it is straightforward to show that  $\overline{D}=\overline{W}=\overline{\dot{\Phi}^+} + \overline{\dot{\Phi}^- }> 0$, which summarizes the Lorenz \nocite{Lor67} (1967) energy cycle \nocite{Lor67}. 
Therefore the atmosphere can be interpreted  as a heat engine, in which $\dot{\Phi}^+$ and $\dot{\Phi}^-$ are the net heat gain and loss needed in order to produce mechanical work $\overline{W}$ given by their (algebraic) sum.
The efficiency of the atmospheric  heat engine, i.e. the capability of generating mechanical work given a certain heat input,   can therefore be defined as: 
\begin{equation}
\eta=(\overline{\dot{\Phi^+ }}  + \overline{\dot{\Phi^-}  })/(\overline{\dot{\Phi^+}})   =  \overline{W}/ \overline{\dot{\Phi}^+}.
\label{eta}
\end{equation}
The analogy between the atmosphere and a (Carnot) heat engine can be pushed further if we introduce the total rate of entropy change of the system, $\dot{S}=\int_V \dot{Q}/T \rho dV$ and use the $V^\pm$ partition of the domain:
\begin{equation}
\hspace*{-0.5 cm} \dot{S}=\int_{V^+} \frac{\dot{Q}^+}{T} \rho dV+ \int_{V^-}  \frac{\dot{Q}^-}{T} \rho dV =  \dot{\Sigma}^+ + \dot{\Sigma}^-  
\label{Srate}
\end{equation}
where $\dot{\Sigma}^+  >0$ and $\dot{\Sigma}^- <0$ at any time. In a  steady state,  $\overline{\dot{\Sigma}^+ }=-\overline{\dot{\Sigma}^- }$ as $\overline{\dot{S}}=0$ and 
 the following expression holds  
\begin{equation}
\frac{\overline{\dot{\Phi}^+ }}{ \Theta^+}  + \frac{\overline{\dot{\Phi}^- }}{ \Theta^-} =0
\label{theta1}
\end{equation}
where   
\begin{equation}
\overline{\dot{\Phi}^\pm} = \overline{\dot{\Sigma}^\pm}   \Theta^\pm
\label{theta2}
\end{equation}
  and the temperatures  $\Theta^+$  and $\Theta^-$  are the time and space averaged temperatures of the $V^+$ and $V^-$ domains respectively. Since $|\dot{\Sigma}^+  |=|\dot{\Sigma}^- |$  and $|\Phi^+|>|\Phi^- |$, it can be shown that $\Theta^+ > \Theta^-$, i.e absorption typically occurs at higher temperature than release of heat \nocite{Peix, Johnson} (Peixoto \& Oort 1992; Johnson 1997).  From Eqs. (\ref{eta}), (\ref{theta1}) and (\ref{theta2}) we derive that $\eta=(\Theta^+ - \Theta^-)/\Theta^+$ \nocite{Johnson2,Lucarini} (Johnson 2000; Lucarini 2009), similarly to the definition of the actual Carnot efficiency \nocite{Fermi} (Fermi 1956) . 
  


In a planet the non-equilibrium steady state is maintained by the global compensation between the net radiative heating in the warm regions (low latitudes on Earth) and net cooling in the cold regions (high latitudes on Earth). The energy budget of the warm and cold regions is closed thanks to the presence of large scale transports performed by the planetary atmosphere. The entropy production due to the irreversibility of the processes occurring within the climatic fluid is called the material entropy production, $\dot{S}_{mat}$ and  can be written in general terms as:
\begin{equation}
\hspace*{-0.5 cm}    \dot{S}_{mat} =    \int_{V} \left( \frac{\epsilon^2}{T}  +  \vec{F} \cdot\nabla\frac{ 1}{T}  \right) \rho dV  
\label{smater}
\end{equation}
where the first term on the right hand side is he contribution coming from the dissipation of kinetic energy, and the second term is related to the transport of heat (in sensible and/or latent heat forms) across gradients of the temperature field. One can prove that the entropy production associated with the dissipation of kinetic energy  is the minimal value of the entropy production compatible with the presence of a Lorenz energy cycle with average intensity $\overline{W}$, 
\begin{equation}
\dot{S}_{min}=\int_V \frac{\epsilon^2}{T}\rho dV
\end{equation}

We can define the irreversibility parameter $\alpha$ measuring the excess of entropy production with respect to the minimum, which results from the heat transport down the temperature gradient:
\begin{equation}
\hspace*{-0.5 cm}   \alpha = \frac{\overline{\dot{S}_{mat}-\dot{S}_{min}}}{\overline{S_{min}}}  = \frac{\int_{V} dV \overline{\vec{F}\cdot\mathbf{\nabla}(\frac{1}{T}) } }{\int_V \overline{ \epsilon^2/T \rho} dV}        
 \end{equation}
 which is zero if all the production of entropy is due to the unavoidable viscous dissipation of the mechanical energy. 
 The parameter $\alpha$ introduced above is related to the Bejan number $\mathcal{B}e$ \nocite{Paoletti} (Paoletti et al. 1989) as $\mathcal{B}e=\alpha+1$.


\section{Model Setup}
\label{Exp}


Bracketing the multi-stability properties of a complex system by resorting to numerous runs of a numerical model with slightly altered values of one or more parameters is indeed a computationally expensive exercise. The matter becomes problematic if we are considering a system as complex as a planetary atmosphere and we are treating a two-dimensional parametric space. In what follows, we focus our attention on an Earth-like planet, and investigate the impact of changing the irradiance $S$ of the parent star, and of changing the opacity of the atmosphere (modulated by the $CO_2$ concentration). Our goal is to find the region in the $(S,[CO_2])$ space supporting multistability for the climate system, and study the transitions between such states. For each point in the $(S,[CO_2])$ space where multistability is found, we refer to the two coexistent states as Warm (W) and Snowball (SB) states. Outside the region of bistability, the climate state will be found in either a W state or a SB state.

The numerical simulations are performed with the Planet Simulator (PLASIM).  PLASIM \cite{Frae2} is a fast-running climate model of intermediate complexity, freely available at \texttt{http://www.mi.uni-hamburg.de/plasim}. 
   Its dynamical core solves the primitive equations for vorticity, divergence, temperature, specific humidity and the logarithm of surface pressure  using the spectral transform method \nocite{Eliasen,Orszag} (Eliasen, Machenhauer \& Rasmussen 1970; Orszag 1970) with semi-lagrangian advection. 
   Interaction between atmosphere and radiation is dealt with simple but realistic  longwave (\cite{Sasamori}) and shortwave (\cite{Lacis}) radiation schemes. The treatment of the solar forcing accounts for both seasonal and the diurnal cycle. 
 Unresolved processes taking place at the subgrid scales as    moist  \nocite{Kuo, Kuo2} (Kuo 1965; 1974) and dry convection, cloud formation and large scale precipitation \nocite{Stephens,Stephens2,SlingoSlingo} (Stephens 1978; Stephens et al. 1982; Slingo \& Slingo 1991), latent and sensible heat boundary layer fluxes, horizontal and vertical diffusion  \nocite{Louis,Louis2,Laursen} (Louis 1979; Louis et al. 1981; Laursen \& Eliasen 1989) are parameterized. The model is coupled to a 50-$m$  slab ocean which contains a thermodynamic sea-ice model. All simulations are performed at axial tilt angle (obliquity) $\delta=23^\circ$. In Table \ref{tab1} there is a summary of  the physical and orbital parameters used in these simulations.
      \begin{table}[ht]
\centering
\caption{List of parameters and symbols. \label{tab1}}
\scalebox{0.77}{
\begin{tabular}{l l ll l  }
\hline
     parameter/symbol                & explanation       &    value \\
                     \hline
         $\Omega_E$            &   Earth's rotation rate  & $7.29\cdot10^{-5} $ rad$^{-1}$\\
      $y$ & orbital year &    $360\times 8.64 \times10^4$ $s$ \\
      $\delta$   &    obliquity    &  $23^\circ $ \\
      $a$ &  planet's radius & $6300$ km \\
   $ c_{d}$    &     specific heat of dry air             &  1004 J\,kg$^{-1}$K$^{-1}$\\
   $ c_{pw}$   &      specific heat of mixed layer model            &   $4180$ J\,kg$^{-1}$K$^{-1}$ \\
   $g$          &  gravitational  acceleration &  $9.81$ m\,s$^{-2}$ \\ 
    $\rho_w$   & ocean water density                   &  $1030$ kg\,m$^{3}$   \\
 $h_{ml}$      &       mixed layer depth           &  $50$ m\\ 
 $p$ & pressure & \\
 $\mathbf v $ & velocity & \\
 $V$ & volume & \\
 $\rho$ & density & \\
   \hline
\end{tabular}
}
\end{table}

The model is run at T21 spectral resolution (approximately $5.6^{\circ}\times 5.6^{\circ}$) with 10 vertical levels. While this resolution is relatively coarse,   it is expected to be sufficient, for slowly rotating and phase locked planets,  for obtaining a reasonable description of the large scale properties of the atmospheric circulations. A higher horizontal resolution is indeed needed at higher rotation rates (e.g. $\Omega \ge 4 \Omega_E$)  since  the size of   baroclinic  structures (Rossby deformation radius, see, e.g. \cite{Holton}) scales as $1/\Omega$.  We remark that previous analyses have shown that  using a spatial resolution approximately equivalent to T21 allows for obtaining an accurate representation of the major large scale features of the climate system and of its global thermodynamical properties \nocite{Jones,Pascale} (Jones et al. 2005; Pascale et al. 2011).

Finally, we consider values of $S$ ranging from 1165 $Wm^{-2}$ to 1510 $Wm^{-2}$, and values of $[CO_2]$ ranging from 90 ppm to 2880 ppm. 

We wish to anticipate a caveat about the appropriateness of the model adopted in this study. The radiation scheme adopted by PLASIM has been devised to deal with Earth-like CO$_2$ concentrations, and in this range it is reasonably accurate. While in this study we do push the radiation code beyond the usually considered conditions, we are still considering CO$_2$ concentrations within one order of magnitude of the present terrestrial conditions, so that we still retain confidence in our model. The effect of considering ultra-high values of CO$_2$ concentrations  has been studied  in detail in (\cite{Pierre05}), where it is discussed under which conditions  using a radiative model designed for Earth climate conditions can be problematic.





\section{Bistability, hysteresis  and regime boundaries in the $[S,CO_2]$ space}
\label{Results}

\begin{figure*}
\centering
 \includegraphics[angle=0, width=0.85\textwidth]{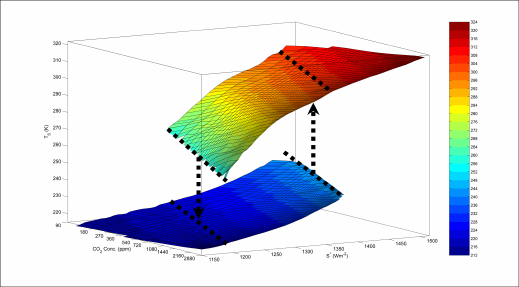}
\caption{ Surface temperature (K) as a function of $S$ and the [CO$_2$]. The lower SB (a) and upper W  (b) manifolds are shown. 
The transition SB$\rightarrow$W and W$\rightarrow$SB are marked with dashed arrows, starting from the $S_{sbw}$ and $S_{wsb}$  tipping regions, respectively.   
               \label{secondo}}
\end{figure*}

\subsection{W and SB states}
\label{manifolds}

In the following we shall  refer  to   the parametric plane ($S$,[CO$_2$])   as the CS space.  The  exploration of the CS space proceeds as follows for each considered value of  $[CO_2]$:
\begin{itemize}
\item[1.]	the model is run to a W steady state for $100$ years with $S$ equal to 1510 W\,m$^{-2}$;
\item[2.]	$S$ is decreased by a small amount and the model run is continued until a steady state is reached;
\item[3.]	step 2 is repeated until $S$ is reduced to 1165 W\,m$^{-2}$; the point of   W$\rightarrow$SB transition  is noted down;
\item[4.]	the reverse operation is then performed with $S$ increased step by step,  up to the value of   1510 W\,m$^{-2}$, each time allowing the system to reach a steady state; the point of SB$\rightarrow$W  transition is noted down.
\end{itemize}

Using systematically this procedure, it is possible to identify two lines in the CS space where the SB$\rightarrow$W and W$\rightarrow$SB transitions occurs. We refer to these  two lines as $S_{sbw}([CO_2])$ and $S_{sbw}([CO_2])$. They have the property that for values of $S$ larger than $S_{sbw}$, only W states are possible, while for values of $S$ smaller than $S_{wsb}$ only SB states are realized. Empirically, one find a rather simple paramerization for the $S_{sbw}$ and $S_{wsb}$ lines: 
\begin{eqnarray}
S_{sbw} = a_{sbw}\log_{10}[CO_2] + C_{sbw}  \label{arr1} \\
   S_{wsb} = a_{wsb}\log_{10} [CO_2] + C_{wsb}     \label{arr2}
\end{eqnarray}
where $a_{sbw}  \approx  a_{wsb} \approx  -70$ Wm$^{-2}$, $C_{sbw} \approx  1630$ Wm$^{-2}$ and $C_{wsb} \approx 1440$ Wm$^{-2}$ for the transition SB$\rightarrow$W and W$\rightarrow$SB respectively and  [CO$_2$] is expressed in ppm.  We can estimate the width $\Delta_B$ of the bistable region as:
\begin{equation}
\Delta_B = C_{sbw}-C_{wsb}\approx    200 \textrm{Wm}^{-2}.
\label{bist_reg}
\end{equation}

The globally averaged mean  surface temperature, $T_s$  in the CS space is illustrated in Fig. \ref{secondo}. Two climatic  regimes can be identified, corresponding to the two leaves of the function $T_s(S,[CO_2])$). The overlap of the two leaves is responsible for the multi stability properties, and the border of the region where both leaves are defined correspond to the tipping points of the system.   $S_{sbw}$ and $S_{wsb}$ are reported as dashed lines in Fig. \ref{secondo}, and the direction of tipping is also indicated.  
Interestingly, in the  CS space,  isotherms are parallel to the $S_{sbw}$ and $S_{wsb}$, so that within a very good degree of approximation the SB$\rightarrow$W occur at $T_s=242$ K, while  the W$\rightarrow$SB take place at $T_s=254$ K.  The bistability properties of the  climate system are therefore ``blind"  to the mechanism of forcing and depend only on the temperature. One must note that surface temperature range $242$ K-$254$ K is not permitted by the climate system.  Across the explored parametric space, the temperature range on the SB and W manifolds are $212$ K-$242$ K, and $254$ K-$326$ K respectively. The climate sensitivity of the W state is much higher than that of the SB state  because the SB is almost entirely dry, so that the positive water vapour feedback is not active and the surface temperature difference between the two manifolds ranges between $40$ K and $ 60$  K, respectively, for identical values of $S$ and [CO$_2$].

\subsection{Transitions and parametrizations}
\label{trans}

There is much more to say than what is portrayed in Fig. \ref{secondo}, \textit{i.e.} the W and SB states differ much more than just in surface temperature. In \cite{Bob} we have presented an extensive account of the thermodynamical properties of the W and SB states in the CS space. An extremely useful fact is that it is possible to establish empirical relations connecting thermodynamical and dynamical properties of a planet to a given thermodynamic quantity,  which can be more easily determined at experimental level, so that we can effectively reduce the 2D CS space to a simpler 1D space. 

Specifically, one can establish approximate empirical laws of the form $\Gamma_{SB}(S,[CO_2])\approx \Gamma^{T_s}_{SB}(T_s(S,[CO_2]))$ and $\Gamma_W(S,[CO_2])\approx\Gamma^{T_s}_W(T_s(S,[CO_2]))$, where $\Gamma$ is a thermodynamical property such as, e.g. entropy production, and the lower index refers to whether we are in the SB or W state. For a given quantity $\Gamma$, the empirical laws $\Gamma^{T_s}_{SB}$ and $\Gamma^{T_s}_{W}$ will in general be different.  This result is quite interesting in a classical perspective of climate dynamics, where it is customary to parameterise large scale climate properties as a function of the surface temperature, especially when constructing simple yet meaningful models (\cite{Saltzman}). This also suggests that, in fact, the surface temperature, well beyond its obvious practical relevance, is, loosely speaking, a good climatic \textit{state variable}, \textit{i.e.} it contains a great deal of information on the physical state of the system.

Similarly, it is possible to establish parametrized laws expressing to a good degree of approximation all the main thermodynamic quantities describing the non-equilibrium steady state of the climate system as a specific functions $\Gamma^{T_E}_{SB}(T_E)$ and $\Gamma^{T_E}_{W}(T_E)$  of the emission temperature $T_E$. This result is even more relevant in terms of both theory and applications. First, we find an entirely non-trivial way to connect non-equilibrium properties to the quantity - $T_E$ - that quintessentially describes the climate system as a zero-dimensional, energy balance system. Second, we have a way to connect a quantity which can be readably observed also in exoplanets - $T_E$ - because only globally integrated energy data are needed - to quantities which are related to horizontal and vertical gradients of the thermodynamic fields of the planetary atmosphere, and thus cannot be directly observed. 

\begin{figure}
\centering
   \subfigure{
    \includegraphics[angle=-0, width=0.45\textwidth]{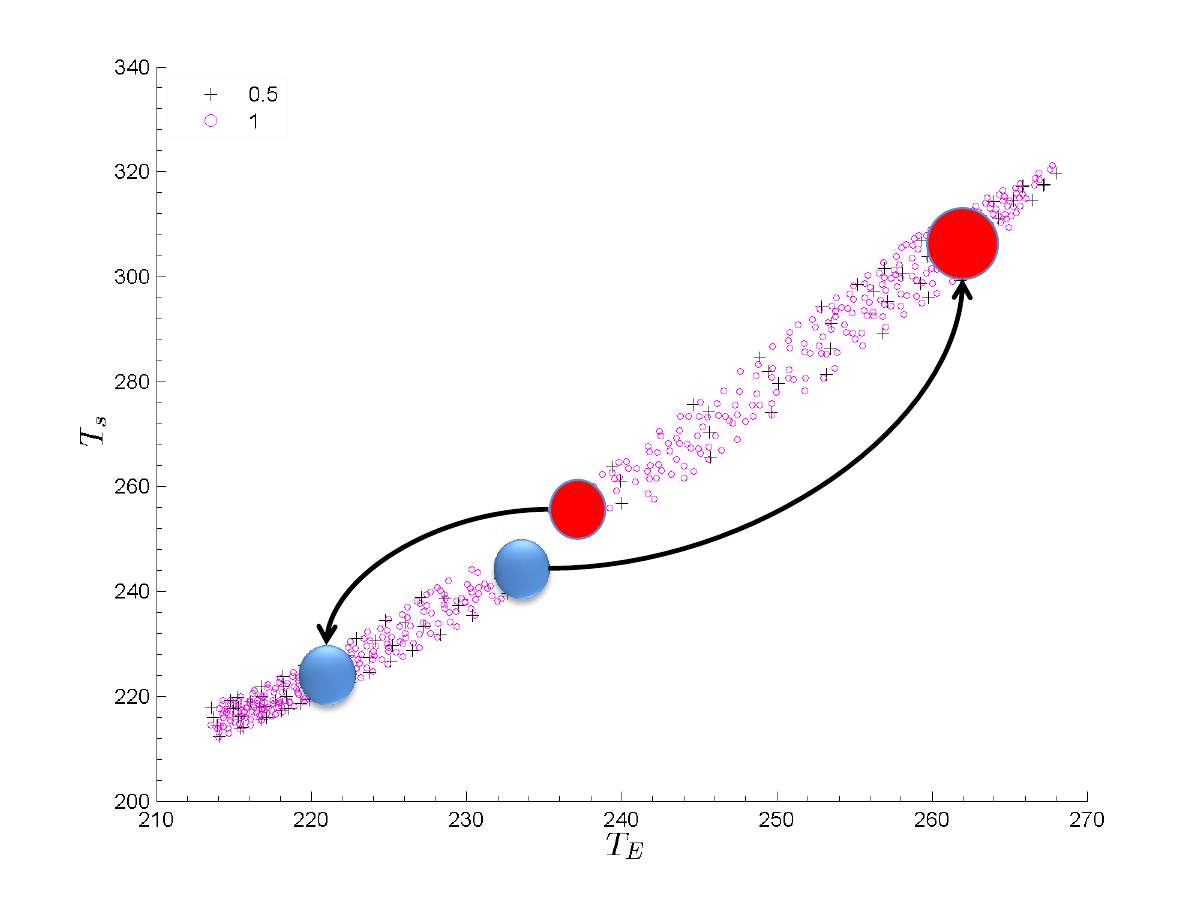}
     } 
\caption{Relationship between emission temperature $T_E$ (K) and surface temperature $T_s$ (K) for $\Omega^*=1$ (magenta) and $\Omega^*=0.5$ (black).    
 \label{tets}}
\end{figure}

We will discuss below these empirical parametrizations. Of course, one may well wonder whether the existence of such parametrizations is robust  and, going more into detail, whether the empirically obtained functional relations are robust with respect to changes in the external parameters of the system, like the orbital ones. Of course, one cannot perform a complete study, but results are extremely encouraging at least considering a very important orbital parameter, the rotation rate of the planet $\Omega$. In Fig. \ref{tets} we present a scatter plot of the emission temperature $T_E$ versus the globally averaged surface temperature $T_s$ for $\Omega^*=1$ (magenta) and $\Omega^*=0.5$ (black), where $\Omega^*=\Omega/\Omega_E$, with $\Omega_E$ the current terrestrial rotation rate. Such a range is much larger than anything ever experienced by our planet, but still far able to encompass the case of slowly rotating planets. We will deal with this case in Sec. \ref{rotation}.  

Figure \ref{tets}, shows that there is a clear monotonic relation between the two temperature indicators, as one could have guessed, and that the $242K-254K$ gap in the allowed values of $T_s$ discussed in Fig. \ref{secondo} corresponds to the gap $233K-238K$ in $T_E$. Moreover, the (approximate) functional relation between $T_s$ and $T_E$ as well as the gaps do not depend substantially on the choice of the rotation rate of the planet. In this figure we also introduce a simplified way for representing the tipping point regions. The SB$\rightarrow$W transitions are represented by the arrow pointing from the blue dot to the red dot, while the W$\rightarrow$SB transitions are represented by the arrow connecting the red to the blue dot. As we see, the projection of the CS space into a 1D space is successful also for the critical states of the system. 

\begin{figure}
\centering
    \includegraphics[angle=-0, width=0.45\textwidth]{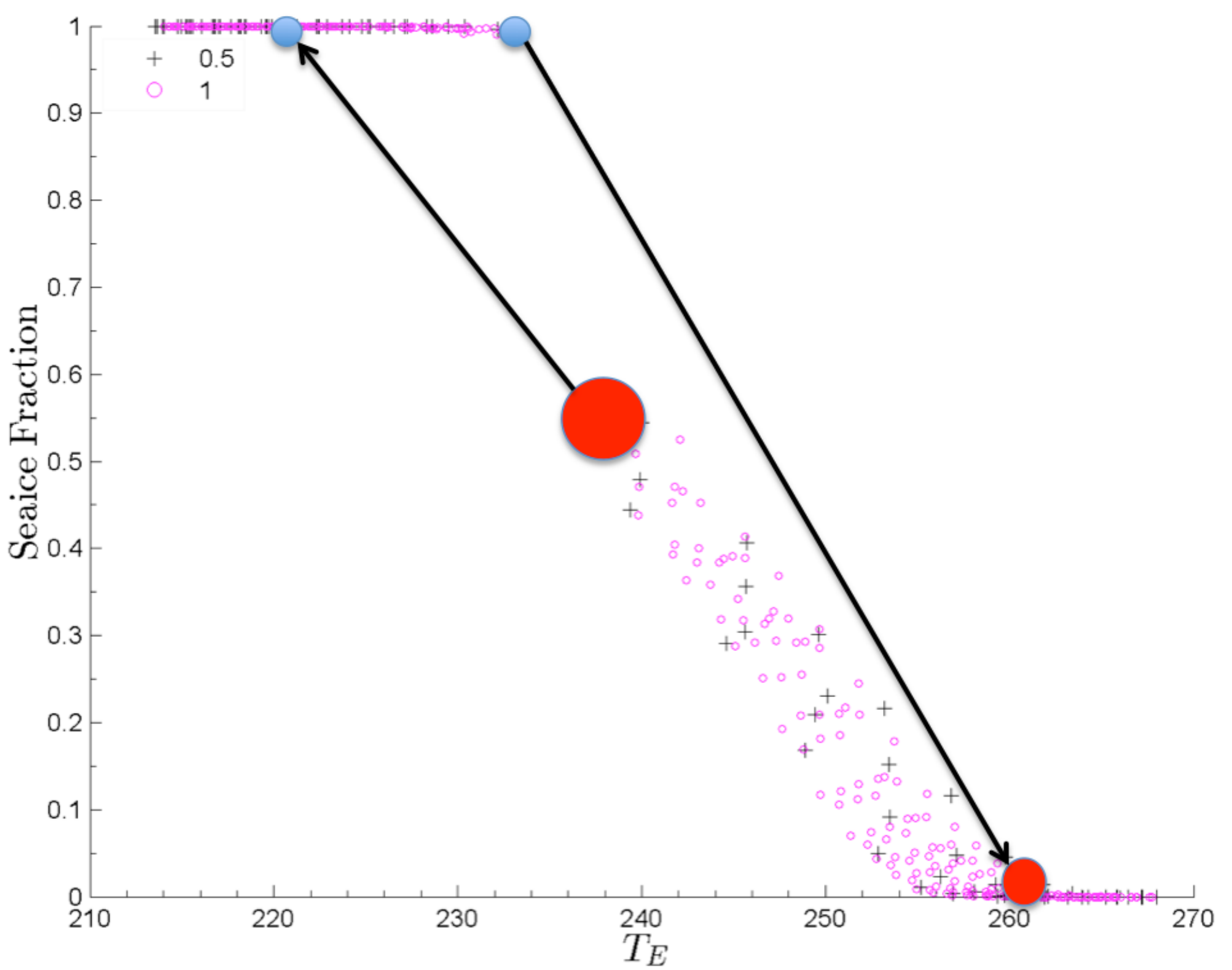}
 \caption{Sea ice fraction (in $\%$)   \emph{vs.}    surface temperature $T_s$ (K) for $\Omega^*=1$ (magenta) and $\Omega^*=0.5$ (black).  
}
 \label{ventuno}
 \end{figure}

Figure \ref{ventuno} shows that the emission temperature is a good predictor of the sea-ice fraction. The SB state, defined by $T_E\leq 233$ K, features an ocean where the sea-ice fraction is unity. The W state has a less trivial behavior: we can identify a subset of states, characterized by $T_E\geq260$ K, where sea ice is absent, and, therefore, the ice-albedo feedbacks is switched off. For $238$ K $\leq T_E\leq260$ K, there is a roughly linear monotonically decreasing relation between the sea-ice fraction and $T_E$. It is interesting to note that  transition from the W to SB states occurs at a critical value of the sea-ice fraction of about 0.5, in broad agreement with previous results by \cite{Voigt2}. The freezing of the planet leads to the sea-ice fraction to unity, whereas the reserved SB$\rightarrow$W transition has even more dramatic effects, because we go from a completely frozen to a virtually ice free ocean. This elucidates even more clearly the strength of the ice-albedo feedback. As we see, there is barely detectable difference between the simulations referring to $\Omega^*=1$ and those referring to  $\Omega^*=0.5$.


\begin{figure}
\centering
    \includegraphics[angle=-0, width=0.45\textwidth]{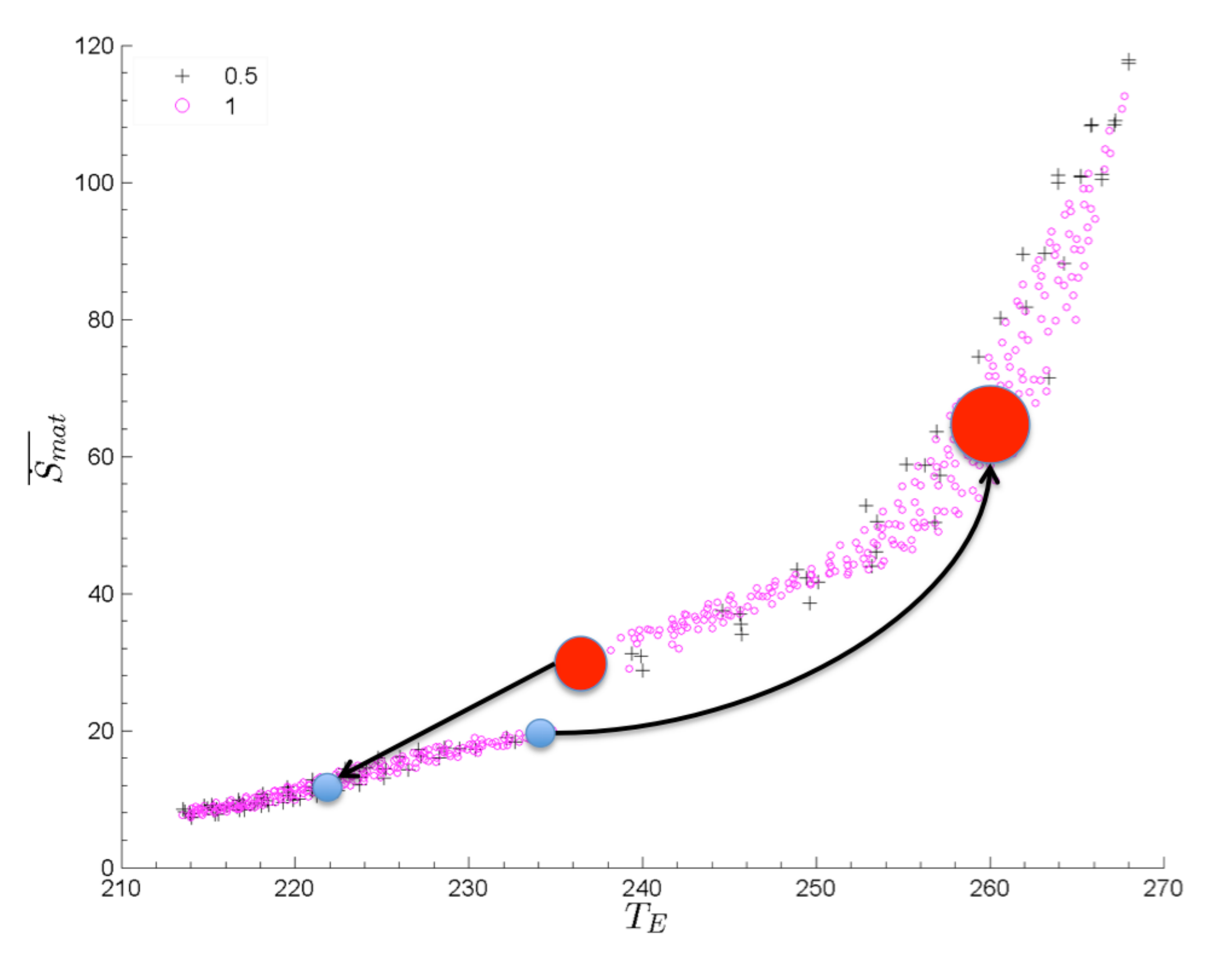}
\caption{Material entropy production $\overline{\dot{S}_{mat}}$ ($10^{-3}$W m$^{-2}$ K$^{-1}$)  \emph{vs.}   emission temperature $T_E$ (K) for $\Omega^*=1$ (magenta) and $\Omega^*=0.5$ (black).   } 
 \label{diciotto}
\end{figure}

\begin{figure}
\centering
    \includegraphics[angle=-0, width=0.45\textwidth]{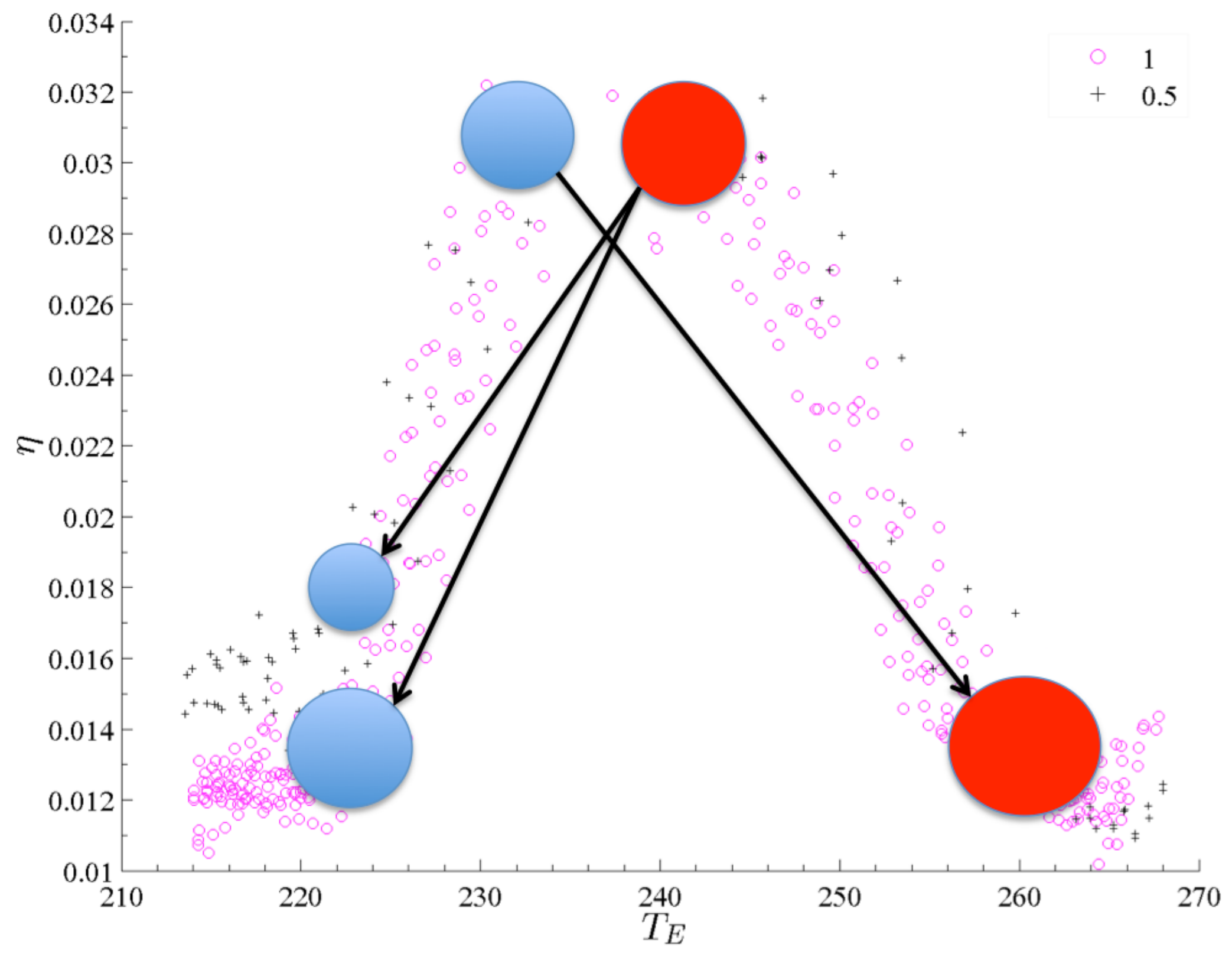}
\caption{Carnot efficiency  \emph{vs.}  emission temperature $T_E$ (K) for $\Omega^*=1$ (magenta) and $\Omega^*=0.5$ (black).}
 \label{diciannove}
 \end{figure}

\begin{figure}
\centering
    \includegraphics[angle=-0, width=0.45\textwidth]{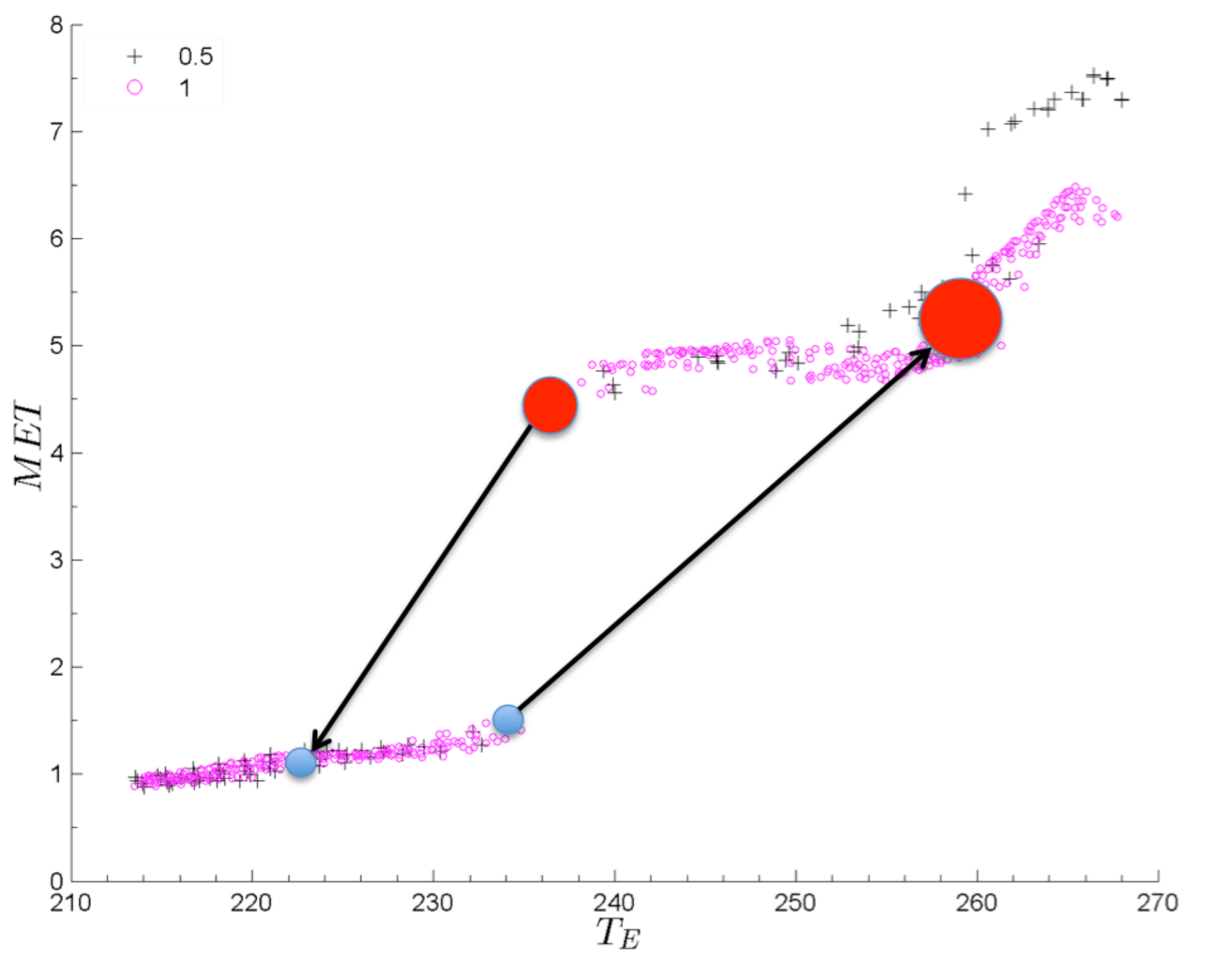}
\caption{Meridional heat transport index $MET$ (in PW) \emph{vs.}   emission temperature $T_E$ (K) for $\Omega^*=1$ (magenta) and $\Omega^*=0.5$ (black).  
 \label{diciassette}}
\end{figure}

In Figs. \ref{diciotto}-\ref{diciassette} we present much less trivial empirical thermodynamic relations, by expressing thermodynamic quantities as approximate functions of the emission temperature $T_E$. As can be guessed by comparing with Fig. \ref{tets}, exactly the same conclusions can be drawn by plotting the same quantities against $T_s$, which actually provides cleaner functional relations \cite{Bob}; nonetheless, we choose $T_E$ as predictor variable because of its astrophysical relevance. 

Figure \ref{diciotto} shows that one can robustly predict the value of the material entropy production, $\overline{\dot{S}_{mat}}$ given the value of $T_E$. We also find that the approximate functional relation is weakly dependent on the rotation rate, as the clouds of black and magenta dots are closely entwined both in the W and SB states.  As a reference, one may consider that in the present climate, the Earth's material entropy productions is about $50\times10^{-3}Wm^{2}K^{-1}$ \nocite{Pascale,Bob} (Pascale et al. 2011; Boschi et al. 2013). The values of $\overline{\dot{S}_{mat}}$ are monotonically increasing with $T_E$. In the SB state, the entropy is exclusively generated by dissipation of kinetic energy and by irreversible sensible heat transport, because the planet is almost entirely dry. As for the W manifold, the main contribution to entropy production comes from latent heat due to large scale and convective precipitation. The presence or lack of a significant hydrological cycle changes entirely the entropy budget of the planet. In the bistable region, the range of $\overline{\dot{S}_{mat}}$  is  $(10 , 19)$ $10^{-3}$W m$^{-2}$ K$^{-1}$   and $(34 , 62)$ $10^{-3}$W m$^{-2}$K$^{-1}$  for the SB and W respectively, therefore a factor of $3$ larger in the W regime with  respect  to the SB regime. Moreover, there is a range of values of $\overline{\dot{S}_{mat}}$  -- from $19$ to $34$ $10^{-3}$W m$^{-2}$ K$^{-1}$ -- which is not allowed by the system. The presence of a large gap and the very pronounced difference between the values of $\overline{\dot{S}_{mat}}$ in the W and SB states  confirms that $\overline{\dot{S}_{mat}}$  may be a better indicator than temperature for discriminating the SB and W states as already discussed in \cite{LucFr}, and is of relevance also regarding the definition of habitability conditions, because large values of $\overline{\dot{S}_{mat}}$ provide strong indication of the presence of an active hydrological cycle.

The radical difference between the W and SB states becomes more apparent when looking into measure of the ability of system to produce mechanical work starting from available potential energy. The Carnot efficiency of the system $\eta$ (Fig. \ref{diciannove}) decreases abruptly with $T_E$ in the W manifold, thus implying that warmer climates  are characterized by smaller temperature differences. The main reason for this is that the transport of water vapour acts as very efficient means for homogeneizing the temperature across the system \nocite{Peix2,LucariniRagone,LucFr} (Peixoto \& Orrt 1992; Lucarini \& Ragone 2011; Lucarini et al. 2011). This matches the fact that the system is characterised by very strong irreversible processes, as described by the very large values of entropy production realized in these conditions (Fig. \ref{diciotto}). Consequently, the value of the parameter of irreversibility $\alpha$  increases as conditions becomes warmer and warmer (not shown); see also \cite{LucFr}. When looking at the intensity of the Lorenz energy cycle $\overline{W}$ (Fig. \ref{venti}), we discover that it is weakly dependent on $T_E$ (but monotonically decreasing with $T_E$ around the present climate) for $\Omega^*=1$, because the strongly reduced efficiency is compensated by the stronger absorption due to the water vapor feedback.  The $\Omega^*=1$ and $\Omega^*=0.5$ cases do not feature a quantitative agreement comparable to what shown in previous indicators, especially regarding the intensity of the Lorenz energy cycle $\overline{W}$, as this is directly related to the dynamical regime of a planetary atmosphere, in turn strongly influenced by $\Omega$ \cite{Pascale3}.
Nonetheless, the qualitative agreement is rather good.  The only detectable difference is that in the W state, the value of $\overline{W}$ is more clearly positively correlated with $T_E$ in the case of slower rotation. The likely reason for the presence of a stiffer climate for $\Omega^*=1$ is that baroclinic instability, more relevant for faster rotation, is very strongly damped by efficient mixing of temperature. 

When considering the SB state, we observe a  strengthening of the dynamics of the climate system with increasing surface (or emission) temperature. This is evident when looking at the dependence of  the strength of the Lorenz energy cycle $\overline{W}$ (Fig. \ref{venti}) and of the Carnot efficiency of the system $\eta$ (Fig. \ref{diciannove}) on the emission temperature . The value of $\overline{W}$ reaches its largest value just at the warm boundary of the SB manifold. Similarly, $\eta$ increases monotonically with increasing temperature. This due to the fact that  in the SB state warmer climates correspond to atmospheric conditions with reduced stratification \cite{Bob}, which allows the development of stronger large scale atmospheric motions. The agreement is rather good for the $\Omega^*=1$ and $\Omega^*=0.5$ cases.


As has been seen, the W$\rightarrow$SB transition is associated to a large decrease of $T_s$, $T_E$,  and  $\dot{S}_{mat}$. The large drop in  $T_s$  is related to the fact that most of the material entropy production is associated with the hydrological cycle that,  in the quasi-dry SB states, is  almost absent. Nonetheless, this does not say much about the processes leading to the transition between the two states, or better, describing how one of the attractors disappears. 
The efficiency $\eta$ (Fig. \ref{diciannove}) features, instead, a more interesting behavior, with its maximum values just before the W$\rightarrow$SB transition.  We find that each transition is associated to a notable decrease (more than 30$\%$) of the efficiency of the system, and the closer the system gets to the transition in the CS space, the larger is the value of the efficiency (in Fig. \ref{diciannove} we can see $\eta$ vs. $T_s$ for $\Omega^*=1$ and $\Omega^*=0.5$). This can be interpreted as follows. If the system approaches a bifurcation point, its positive feedbacks become relatively stronger and the negative feedbacks, which act as re-equilibrating mechanisms, become less efficient. As a result, the differential heating driving the climate is damped less effectively, and the system is further from equilibrium, since larger temperature differences are present. At the bifurcation point, the positive feedbacks prevail and the circulation, even if rather strong, is not able to cope with the destabilizing processes, and the transition to a state on the other  manifold is realized. The new state is  more stable  and  closer to thermodynamical  equilibrium. The decreased value of the efficiency looks to be  the marker of this property.  It is interesting to note that, at the transitions,  the values of the efficiency are practically equal:   $\eta_{W\rightarrow SB}\approx \eta_{SB,W}\approx 0.03$. Moreover, we observe that $\eta$ saturates in the very cold regime of SB states and very warm regime of W (Fig.~\ref{diciannove})  at $\eta_{sat,W}\approx \eta_{sat,SB}\approx 0.012$.  This leaves us with an open question: is the result on the decrease of the efficiency at the bifurcation points in both W$\rightarrow$SB and SB$\rightarrow$W transitions of more general relevance?

\begin{figure}
\centering
    \includegraphics[angle=-0, width=0.45\textwidth]{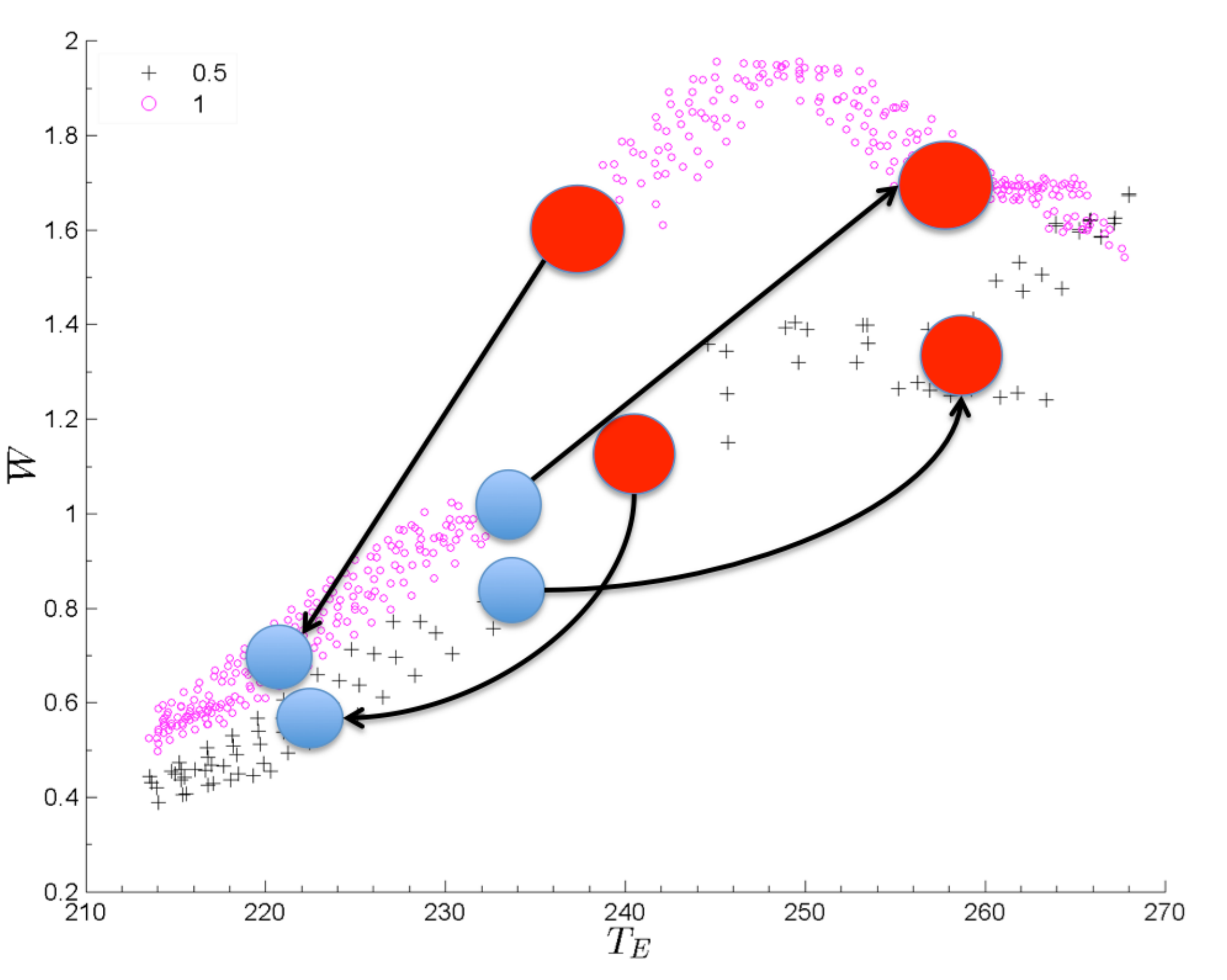}
    \label{venti}
\caption{Lorenz energy cycle strength $\overline{W}$ (W m$^{-2}$)    \emph{vs.} emission temperature $T_E$ (K) for $\Omega^*=1$ (magenta) and $\Omega^*=0.5$ (black)  
 \label{venti}}
\end{figure}

As last element of the non-equilibrium properties of the climate system, we discuss the relationship between $T_E$ and the meridional energy transport $MET$ (Fig. \ref{diciassette}). We observe that in the SB regime there is a weakly positive relationship between $T_E$ and the transport. As discussed above, this is related to the fact that warmer conditions allow for a reduced static stability of the atmosphere. Instead, for a vast range of values of $T_E$ in the W regime, the transport is almost insensitive to $T_E$ (and $T_s$), as discussed thoroughly in \nocite{Caballero05,Battisti1,Battisti2} (Caballero \& Langen 2005; Donohoe \& Battisti 2011; 2012). The rigidity of the climate system has been attributed by \nocite{Battisti1,Battisti2} Donohoe \& Battisti (2011; 2012) to  the fact that the meridional heat transport is mainly determined by planetary albedo and thus atmospheric composition rather that surface albedo and sea-ice coverage.  However, such a mechanism ceases to exist when the ice albedo feedback becomes ineffective because of the disappearance of sea-ice above a threshold value of $T_E\approx 260$ K. Above this value, we observe a steep monotonic increase of the transport with temperature, because  changes in the latent heat transport are mainly responsible for this behavior. This agrees with the idea that the dynamics and sensitivity of a warm planet is, in some sense, dominated by the hydrological cycle. In this latter case, the skill of the two temperature quantities in parameterising the $MHT$ is comparable. Indeed, the details of the transfer functions will depend on various specific properties of the planetary system under investigation. The focus here is on the fact that our results suggests that it is possible to define such empirical relations. As in most previous cases, there is overall a very good agreement between the $\Omega^*=1$ and $\Omega^*=0.5$ cases; the stronger positive correlation between $T_E$ and $MHT$ in the W state can be interpreted similarly to what done when looking at $\overline{W}$, since a stronger atmospheric circulation tends to have a more intense $MHT$.


\section{How does rotation affect bistability?}
\label{rotation}

\begin{figure*}
\centering
\subfigure[]{
 \includegraphics[angle=-0, width=0.45\textwidth]{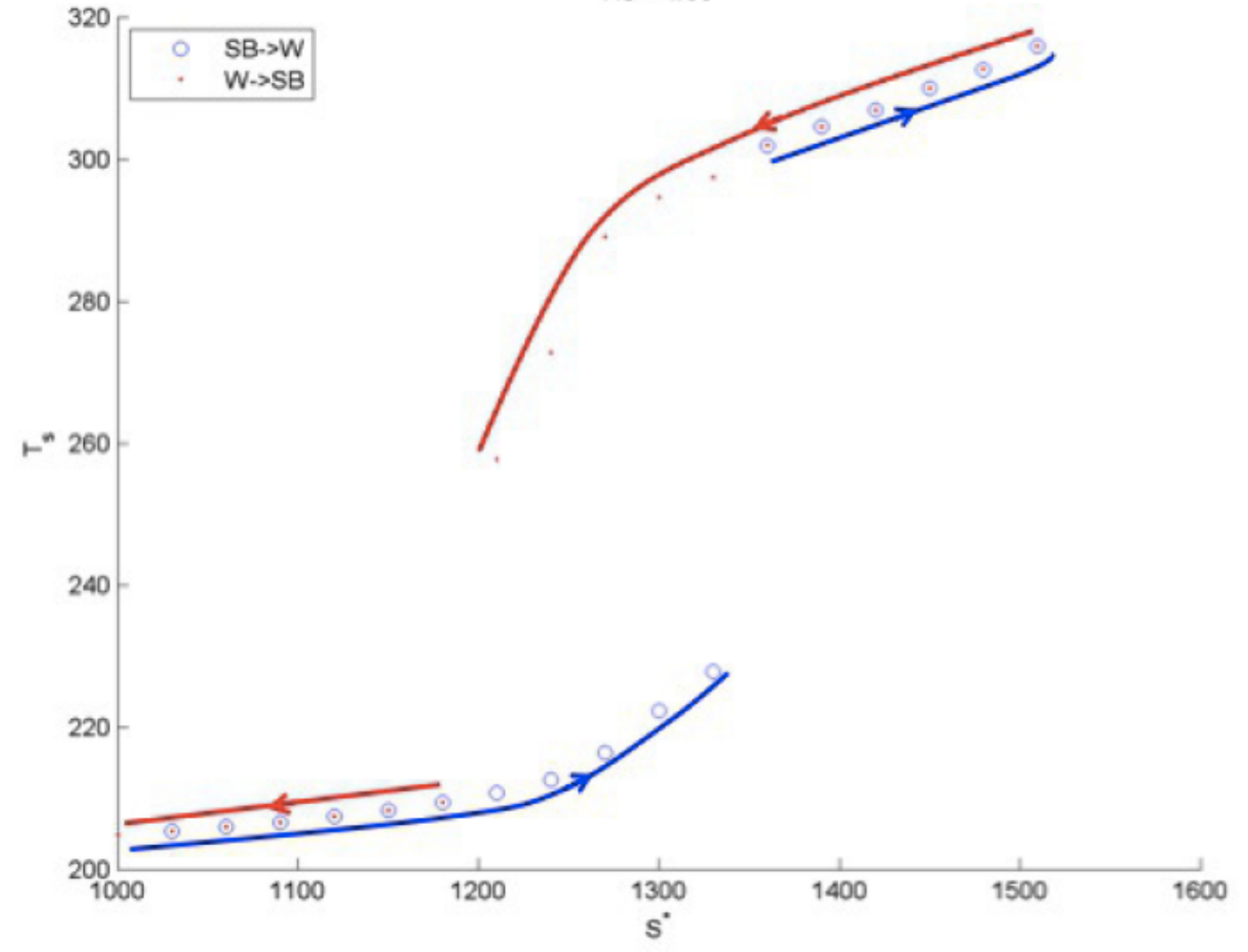}
  \label{ventidue_a}
   }
   \subfigure[]{
    \includegraphics[angle=-0, width=0.45\textwidth]{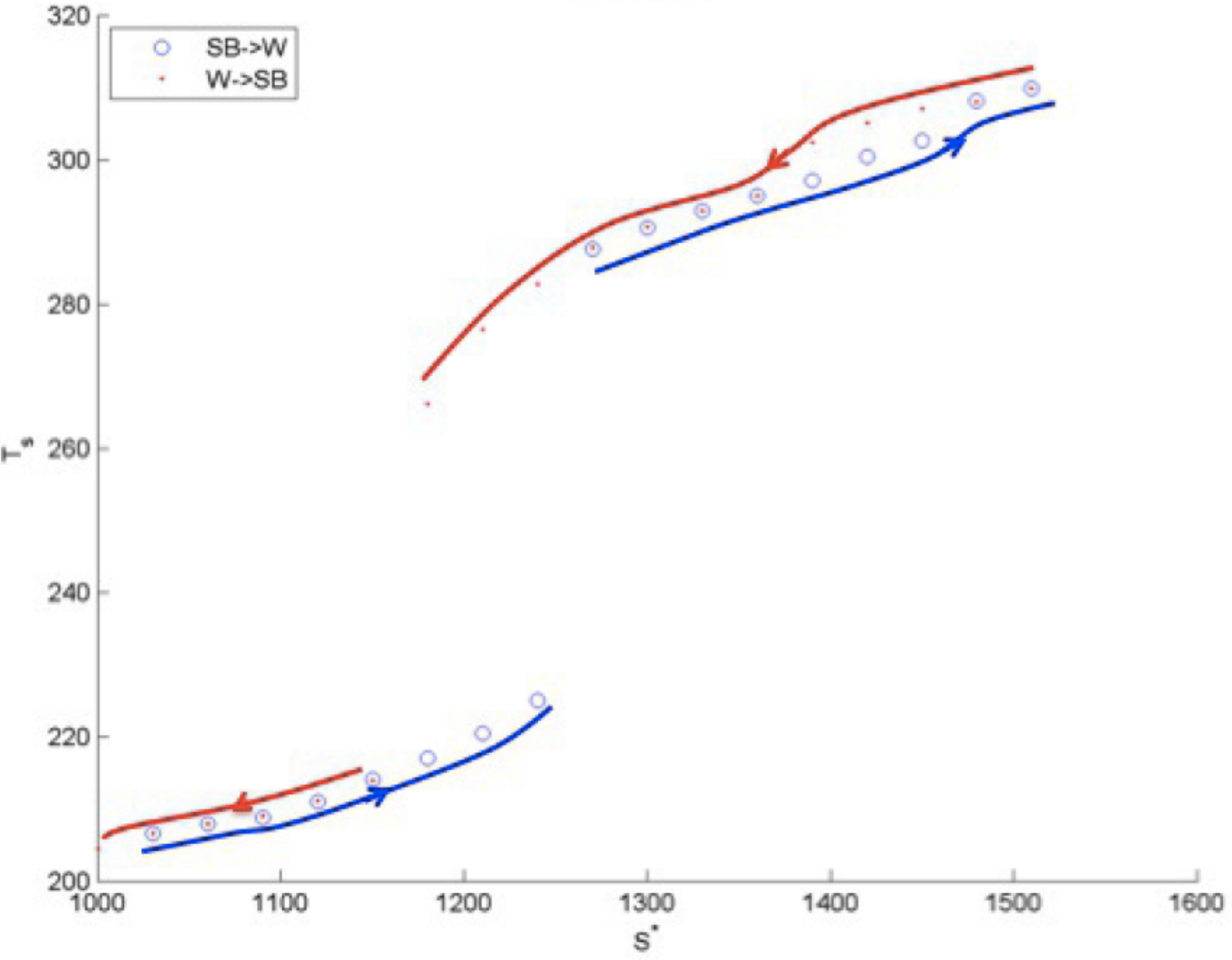}
    \label{ventidue_b}
     } 
   \subfigure[]{
    \includegraphics[angle=-0, width=0.45\textwidth]{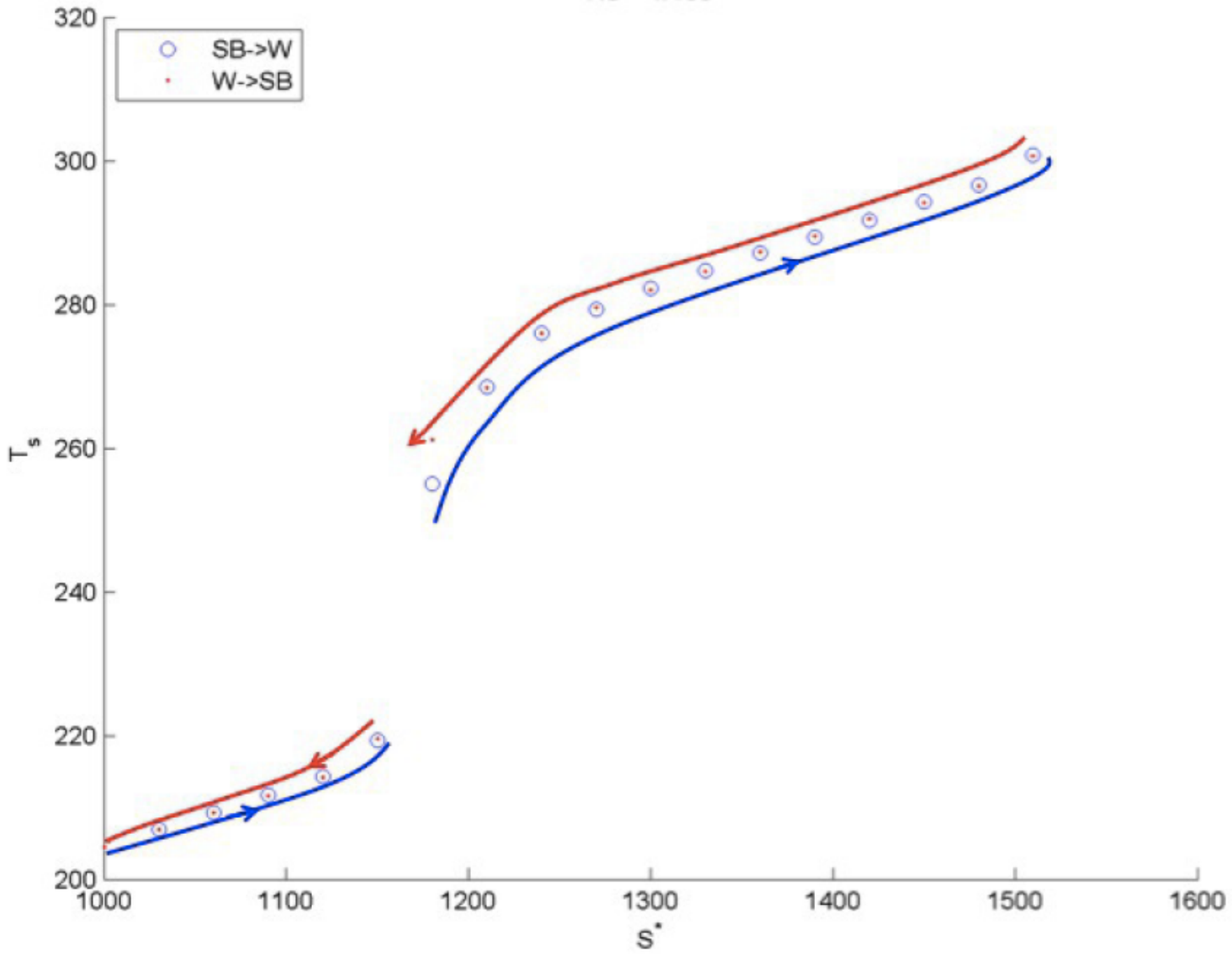}
    \label{ventidue_c}
     } 
   \subfigure[]{
    \includegraphics[angle=-0, width=0.45\textwidth]{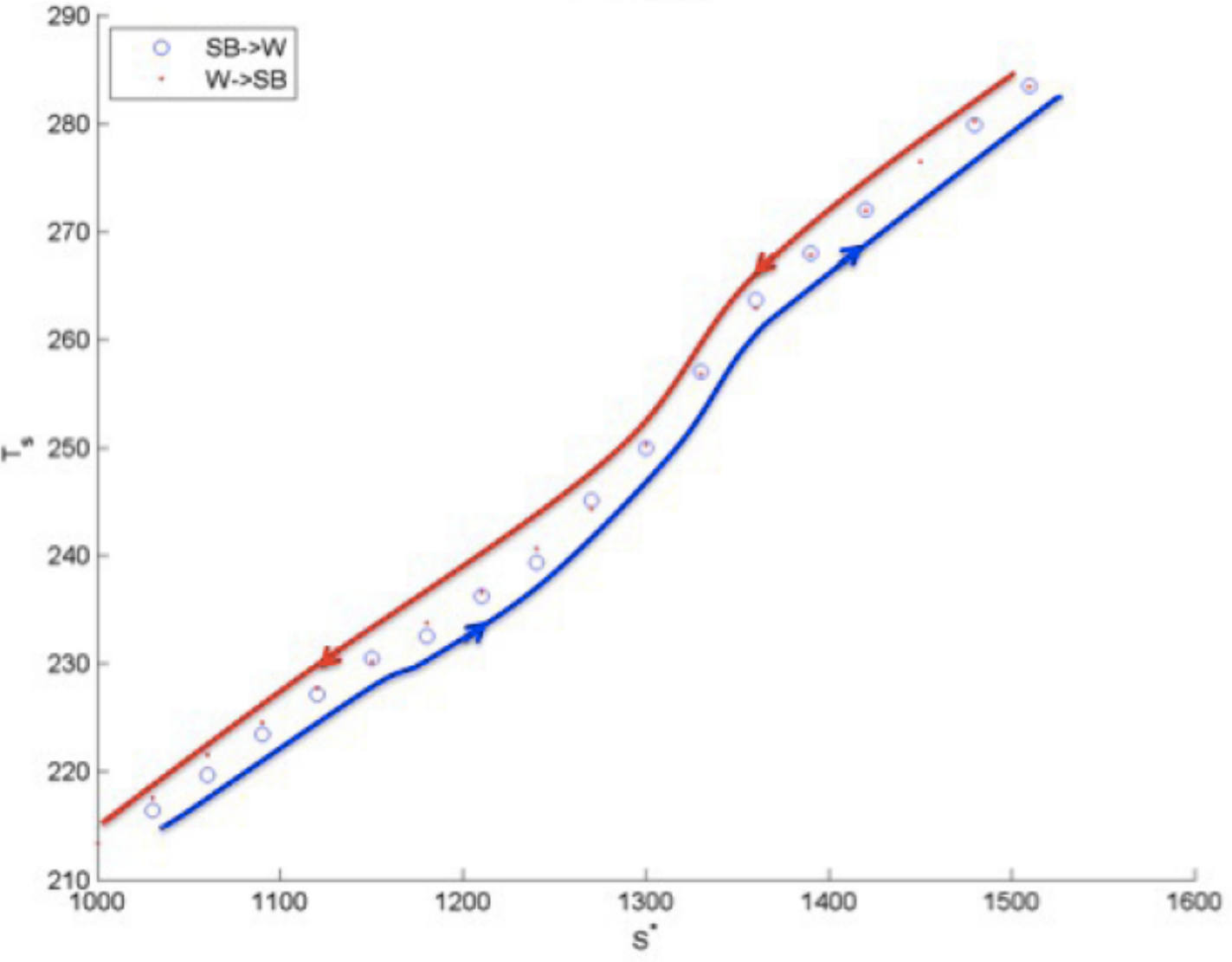}
    \label{ventidue_d}
     } 
\caption{Surface temperature $T_s$ (K) \emph{vs} solar constant $S$ ($Wm^{-2}$) for (a) $\Omega^*=1/30$, (b) $\Omega^*=1/120$, (c) $\Omega^*=1/180$ and $\Omega^*=1/360$. 
 \label{ventidue}}
\end{figure*}

The results presented so far   have been obtained making specific choices  such as the position and size of the continents, the radius of the planet, the eccentricity of its orbit, the orbital tilt and its rotation rate, the atmospheric composition, among others. Obviously, it is hard to account for  such a vast multidimensional parametric space, and one must come to terms with the fact that only partial explorations are possible (and future investigations will deal with different axes of such parametric space). 
A first parameter which is  relevant in determining  the dynamical and thermodynamical properties of the atmospheric general circulation is the rotation rate of the planet \nocite{Williams_a, Williams_b,  Hunt, Pascale3} (Williams 1998a; 19988b; Hunt 1979; Pascale et al. 2013 ). In the previous section,  we considered  two values of the rotation rate,  $\Omega_E$ and $\Omega_E/2$. Our results have clarified that only minor differences exist between the  two cases. However, the explored range falls short of exploring all possible or at least all observed planetary conditions. In particular,  many exoplanets for which habitability conditions are debated are phase-locked as a result of tidal forces, so that they always face the parent star  with the same  hemisphere \nocite{Dvorak,Seager, Kasting, Perryman} (Dvorak 2008; Kastings 2009; Seager 2010; Perryman 2011), and their atmospheric circulation is driven by the temperature difference between the bright, warm side (facing the parent star), which receives radiation, has a positive radiative budget at the top of the atmosphere, and redistributes heat towards the dark, cold side, which emits longwave radiation to space (\cite{Merlis}). 

Such a forcing/dissipation scheme is conceptually similar with respect to the case of fast rotating planets like Earth, with the crucial difference that the boundary conditions are basically time-independent. Atmospheres and circulations of phase-locked planets are currently intensively investigated (\cite{Joshi}; \cite{Heng1};  \cite{Heng2}; \cite{Merlis}). In general, it is reasonable to expect that phase-locked and, more generally, slowly rotating  planets feature very different dynamical and  thermodynamical properties, for the basic reason that the scaling of the underlying evolution equation is entirely different from the case of fast rotating planets. This causes technical problems when attempting to adapt a model devised for describing terrestrial or Earth-like conditions for describing the circulation of slowly rotating planets. In our case, we have extensively rewritten the PLASIM's code \cite{Frae2} so that, in a rather unusual way in the context of climate modeling, the evolution equations are not scaled with respect to given constants. This is the first step towards the construction of a fully flexible modeling suite for super-Earths.  

We investigate the bistability of a slowly rotating Earth-like planet by using a modified  model set-up, where, for sake of generally, we choose the case of zero eccentricity, we remove the land-sea contrast typical of our planet, and an ocean-only surface is considered (so-called \textit{Aquaplanet} set-up). We perform simulations considering the case of extremely low rotation ($\Omega^*=1/180, 1/120, 1/90, 1/30$) plus the case of 1:1 phase lock condition. This corresponds to selecting $\Omega^*=1/360$; note that, as shown in Table \ref{tab1}, the length of the orbital year is chosen to be equal to 360 usual terrestrial days. For each value of $\Omega^*$, the $CO_2$ concentration is kept at $360$ ppm, while the solar constant $S$ is slowly decreased from $1500$ $Wm^{-2}$ to $1000$ $Wm^{-2}$, and increased again from $1000$ $Wm^{-2}$ to $1500$ $Wm^{-2}$.  The parametric exploration follows the same steps as described in Section \ref{Exp} and allows for detecting hystereses in the climatic variables, which can be related to the multi-stability of the system.
 
Since in this case the goal is to detect potential variations in the structural properties of the climate system resulting from changes in $\Omega^*$, we focus on a single climatic variables. In  Fig. \ref{ventidue} (a)-(d) we show the dependence of the globally averaged surface temperature $T_s$ as function of the solar constant $S$ for some of the considered values of $\Omega^*$.  The red line marks the climatic states obtained when $S$ is decreased $1500$ $Wm^{-2}$ $\rightarrow$  $1000$ $Wm^{-2}$, whereas the blue line describes the states obtained when $S$ is increased $1000$ $Wm^{-2}$ $\rightarrow$  $1500$ $Wm^{-2}$.  When comparing the sets of runs performed with $\Omega^*=1/30$ - Fig. \ref{ventidue_a} - and  $\Omega^*=1/120$ - Fig. \ref{ventidue_b} -, we observe that the values of $T_s$ in the W and SB states are rather similar. An important difference emerges, though: the width of bistability  $\Delta_B$ (compare definition in Eq. \ref{bist_reg}) is greatly reduced for a more slowly rotating planet, going from $\sim 150Wm^{-2}$ to $\sim 90Wm^{-2}$.  The amplitude of the bistability interval of $S$ is further reduced by decreasing $\Omega^*$, up to $\Omega^*=1/180$, which is the largest value of $\Omega^*$ for which $\Delta_B\sim0$ W\,m$^{-2}$ (Fig. \ref{ventidue_c}). Correspondingly, in this case, near the tipping points the slope of $T_s$ as a function of $S$ tends to infinity both on the left and on the side of the critical value $S_{crit}\sim1170$ $Wm^{-2}$. 

\begin{figure}
\centering
   \subfigure{
    \includegraphics[angle=-0, width=0.4\textwidth]{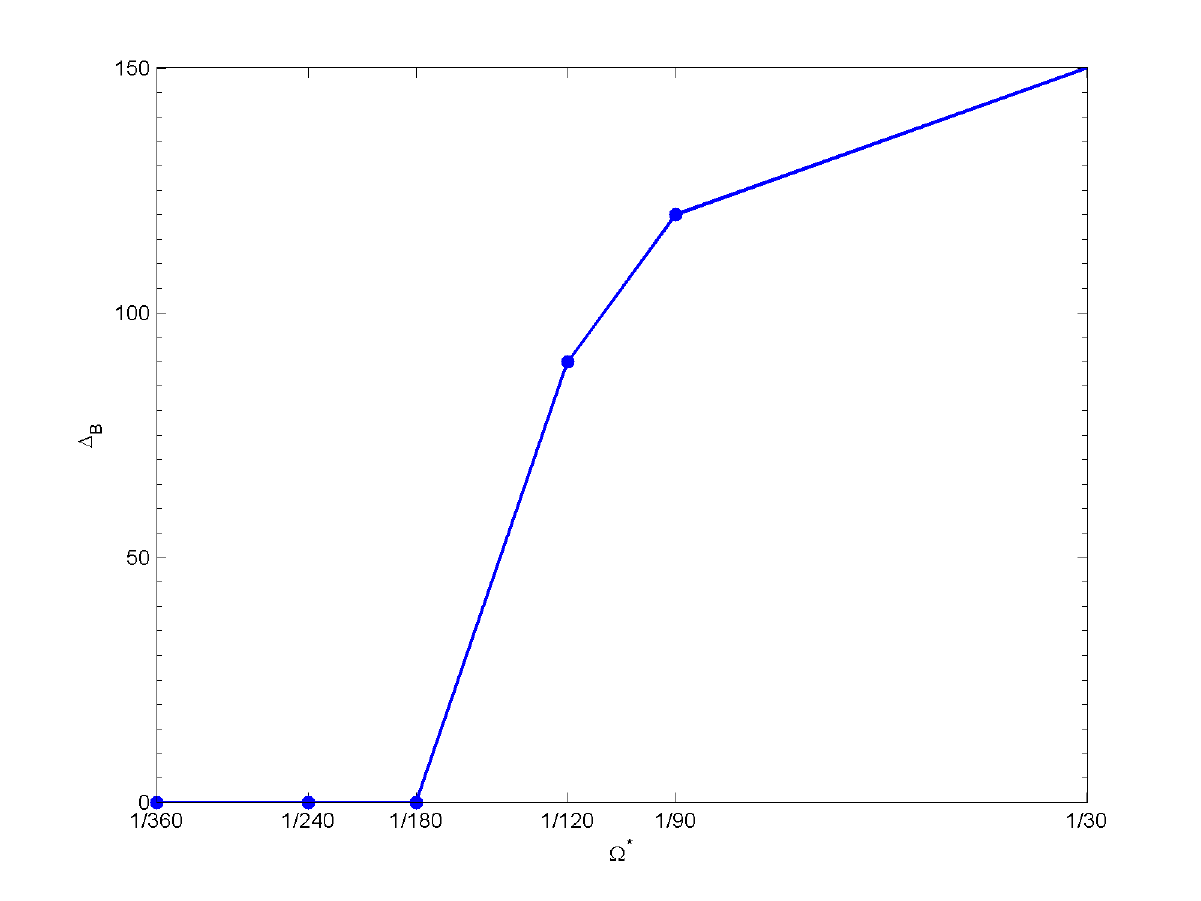}
     } 
\caption{Bifurcation graph of the planetary circulation: width of the bistable region $\Delta_B$ ($Wm^{-2}$) \emph{vs}. $\Omega^*$.  
 \label{bist}}
\end{figure}

Hence, we can introduce $\Omega^{*}_{crit}\sim1/180$ as critical value for the parameter $\Omega^*$, defining the occurrence of a phase transition in the planetary system. The bifurcation graph is reported in Fig. \ref{bist}, where the width of the bistability region $\Delta_B$  is plotted against $\Omega^*$. 

For  $\Omega^*\leq\Omega^{*}_{crit}$, monostability is detected and, as expected, no divergence in the derivative of $T_s$ with respect to $S$ is found, while a rather regular monotonic dependence $\mathrm{d}T_s/\mathrm{d}S>0$ is found. See in Fig. \ref{ventidue_d} the special case of 1:1 phase locked planed ($\Omega^*=1/360$).
In this case,  steady states are characterised by equatorial sea-ice building up in the dark side of the planet and ice-free bright side; looking at the relationship between sea-ice fraction and $S$ (not shown), one finds a very regular monodic dependence, with no threshold effects of the sort seen in Fig. \ref{ventuno}. In the considered range of $S$, the sea-ice fraction is always below unity and above zero. We have that the ice-albedo feedback becomes weaker - to the point that  bistability is lost - when, no matter the intensity of the incoming radiation, the equatorial region spends enough time on the bright side of the planet to allow for melting of sea ice, and, conversely, the polar regions spend enough time in the dark side of the planet to allow for freezing of surface water, unless we consider very low or high values of the constant $S$. We guess that other parameters of great relevance in this context are  the obliquity of the planet and the relationship between the length of the year, which we have kept fixed to the usual one in these simulations, and the time scale describing the rate of relaxation of the temperature of the ocean's surface (\cite{Pascale3}).  

Therefore, when no bistability is present, it is possible to relate one-to-one the solar constant to the globally averaged surface temperature (or to the emission temperature) and no gap in the possible values of the globally surface temperature is found. In other words, once we lose the global instability related to the ice-albedo feedback, the system becomes rather boring or,alternatively, much more easy to interpret. Habitability conditions are also  more easy to assess.

\section{Conclusions}
\label{concl}
In this review   we have examined the bistability properties  of an Earth-like planet using a set of recently developed thermodynamical diagnostics, which allow for defining the fundamental non-equilibrium properties of any planetary atmosphere, ranging from material entropy production, efficiency of the climate engine powering the atmospheric circulation, and linking them to more classical dynamical indicators such as the intensity of the Lorenz energy cycle. 

We have used and modified according to our needs the open-source modeling suite PLASIM, an intermediate complexity global terrestrial climate model, by allowing for an extensive parametric exploration of rather diverse planetary conditions, in terms of amount of incoming stellar radiation, opacity of the atmosphere (modulated by the $CO_2$ concentration), and rotation rate of the planet. 
 
Within  a wide parametric space, which includes the present conditions, the climate is multistable, i.e. there are two coexisting attractors, one characterised by warm conditions, where the presence of sea-ice and seasonal snow cover is limited or altogether absent (W state), and one characterised by a completely frozen sea surface, the so-called snowball (SB) state. As well known, this fact has paleoclimatological relevance, but this is not the main direction of our study.
	
For all considered values of [$CO_2$] (from $90$  to $2880$ ppm) the width of the bistable region is about $200$ Wm$^{-2}$ in terms of the value of the solar constant, and its position depends linearly on the logarithm of the [$CO_2$] and  shifting by about $15$ Wm$^{-2}$ per doubling of $CO_2$ concentration. The W state is characterized by surface temperature being $40$ K - $60$ K higher than in the SB state. In the W states, the material entropy production is larger by a factor of 4 (order of $40-60$ 10$^{-3}$Wm$^{-2}$K$^{-1}$  vs.  $10-15$ 10$^{-3}$Wm$^{-2}$K$^{-1}$ with respect to the corresponding SB states. The boundaries of the bistable region are approximately isolines of the globally averaged surface temperature or of the emission temperature, and in particular, the warm boundary, beyond which the SB state cannot be realized, is characterized by the vanishing of the permanent sea-ice cover in the W regime. 
                 
              The thermodynamical and dynamical properties of the W and SB states are largely different. In the W states,  the hydrological cycle dominates the dynamics and latent heat fluxes contribute most in redistributing the energy in the system and to the generation of material entropy production. The SB state is eminently a dry climate, with entropy production mostly due to sensible heat fluxes and  dissipation of  kinetic energy. The response to increasing temperatures of the two states is rather different: the W states feature a decrease of the efficiency of the climate machine, as enhanced latent heat transports kill energy availability by reducing temperature gradients, while in the SB states the efficiency is increased, because warmer states are associated to lower static stability, which favors large scale atmospheric motions. The entropy production increases for both states, but for different reasons: the system become more irreversible and less efficient in the case of W states, while stronger atmospheric motions lead to stronger dissipation and stronger energy transports in the case of SB states.

A general property which has been  found is that, in both regimes, the efficiency increases for steady states getting   closer to tipping points  and  dramatically drops at the  transition to the new state belonging to the other attractor. In a rather general thermodynamical context, this can be framed as follows: the efficiency gives a measure of how far from equilibrium the system is. The negative feedbacks tend to counteract the differential heating due to the stellar  insolation pattern, thus leading the system closer to equilibrium. At the bifurcation point, the negative feedbacks are overcame by the positive feedbacks, so that the system makes a global transition to a new state, where, in turn, the negative feedbacks are more efficient in stabilizing the system. On a more phenomenological note, the transition from the W to SB states occurs at a critical value of the sea-ice fraction of about 0.5. This agrees with previous findings (Voigt and Pierrehumbert 2011). After the transitions, the sea-ice fraction becomes unitary. When considering the reverse tipping point, the transition is even more dramatic: the sea-ice fraction changes abruptly from unity to virtually zero. 
               
We have shown that empirical functional relations are found between the main thermodynamical quantities  and globally averaged surface temperature of the emission temperature,  allowing for expressing the global non-equilibrium thermodynamical properties of the system in terms of parameters which are more directly accessible.  Although this method requires further investigation in order to delimit its range of applicability, it suggest a methodology to infer information about the atmospheric dynamics of exoplanets (which would be otherwise unaccessible), also because we have discovered that the transfer functions are rather robust with respect to large variations of orbital parameters such as the planetary rotation rate.

In the last part of this work we have explored  the dynamical range of slow rotating and phase locked planets, where the length of the day and the length of the year are comparable. We have clearly found that there is critical rotation rate below which the multi-stability properties are lost, and the ice-albedo feedback responsible for the presence of SB and W conditions is damped. The bifurcation graph of the system suggests the presence of a phase transition in the planetary system. As such, critical rotation rate corresponds roughly to the phase lock 2:1 condition. More specifically, the width of the bistable region is gradually reduced up to zero for  $\Omega^*\approx 1/180$. For $\Omega^*\leq 180$ the two attractors associated with the W state (polar sea-ice caps and tropical ice-free ocean) and the SB state (globally covered by sea-ice) merge and only one attractor exist, corresponding to a totally different climate (equatorial sea-ice in the dark side and ice-free ocean in the bright side). In particular, if an Earth-like planet were phase locked 1:1 with respect to its parent star, only one climatic state would be compatible with a given set of astronomical and astrophysical parameters. We plan to extend the investigation of phase-locked planetary conditions by exploring the impact of changing the length of the orbital year, thus performing a thorough analysis of the thermodynamical properties of the the circulation patterns described by along the lines of \nocite{Merlis} Merlis \& Schneider (2010). 

The results discussed in this paper support the adoption of new diagnostic tools based on non-equilibrium thermodynamics  for  analysing the fundamental properties of planetary atmospheres and pave the way for the possibility of practically deducing fundamental properties of planets in the habitable zone from relatively simple observables. Future investigations will analyse more systematically how robust our findings are with respect to changes in relevant orbital parameters.

\acknowledgements
The authors acknowledge that the research leading to these results
has received funding from the European Research Council
under the European Community's Seventh Framework
Programme (FP7/2007-2013) / ERC Grant agreement
No. 257106, project Thermodynamics of the Climate
System - NAMASTE, and has been supported by the
Cluster of Excellence CLISAP.The authors would like to thank  K. Fraedrich, P. Hauschildt, F. Lunkeit  and F. Ragone for their comments and insightful discussions. VL wishes to thank the German Astronomical Society for the invitation to present these scientific results at the annual assembly held in Hamburg in September 2012.


\end{document}